\documentclass[preprint,review,12pt,sort&compress]{elsarticle}
\usepackage[margin=1.18 in]{geometry}
\usepackage{multirow}
\usepackage[usenames, dvipsnames]{color}
\definecolor{UW}{RGB}{64, 38, 96}

\usepackage{amssymb}

\usepackage{float}

\journal{Carbon}

\begin{document}
\begin{titlepage}

\clearpage\thispagestyle{empty}



\noindent

\hrulefill

\begin{figure}[h!]

\centering

\includegraphics[width=1.5 in]{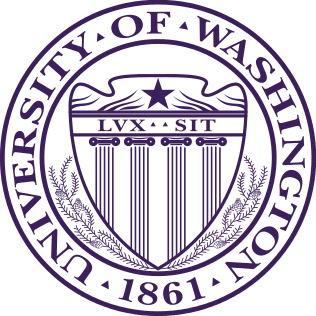}

\end{figure}


\begin{center}

{\color{UW}{

{\bf A\&A Program in Structures} \\ [0.1in]

William E. Boeing Department of Aeronautics and Astronautics \\ [0.1in]

University of Washington \\ [0.1in]

Seattle, Washington 98195, USA

}

}

\end{center} 

\hrulefill \\ \vskip 2mm

\vskip 0.5in

\begin{center}

{\large {\bf Failure and Scaling of Graphene Nanocomposites}}\\[0.5in]

{\large {\sc Cory Hage, Yao Qiao, Marco Salviato}}\\[0.75in]

{\sf \bf INTERNAL REPORT No. 17-02/01E}\\[0.75in]

\end{center}

\noindent {\footnotesize {{\em Submitted to Carbon \hfill February 2017} }}

\end{titlepage}

\newpage

\begin{frontmatter}


\cortext[cor1]{Corresponding Author, \ead{salviato@aa.washington.edu}}

\title{Failure Behavior and Scaling of Graphene Nanocomposites}


\author[address]{Cory Hage Mefford}
\author[address]{Yao Qiao}
\author[address]{Marco Salviato\corref{cor1}}

\address[address]{William E. Boeing Department of Aeronautics and Astronautics, University of Washington, Seattle, Washington 98195, USA}

\begin{abstract}
\linespread{1}\selectfont

This work proposes an investigation on the scaling of the structural strength of polymer/graphene nanocomposites. To this end, fracture tests on geometrically scaled Single Edge Notch Bending (SENB) specimens with varying contents of graphene were conducted to study the effects of nanomodification on the scaling. 

It is shown that, while the strength of the pristine polymer scales according to Linear Elastic Fracture Mechanics (LEFM), this is not the case for nanocomposites, even for very low graphene contents. In fact, small specimens exhibited a more pronounced ductility with limited scaling and a significant deviation from LEFM whereas larger specimens behaved in a more brittle way, with scaling of nominal strength closer to the one predicted by LEFM.

This behavior, due to the significant size of the Fracture Process Zone (FPZ) compared to the specimen size, needs to be taken into serious consideration. In facts, it is shown that, for the specimen sizes investigated in this work, neglecting the non-linear effects of the FPZ can lead to an underestimation of the fracture energy as high as 113\%, this error decreasing for increasing specimen sizes. 

\end{abstract}

\begin{keyword}
Graphene, Fracture, Size effect, Damage



\end{keyword}

\end{frontmatter}


\section{Introduction}
\label{intro}
Thanks to their outstanding specific mechanical and functional properties, the engineering use of polymer/graphene nanocomposites is becoming broader and broader. Current applications include electronics, additive manufacturing, energy storage devices and the use as nanoreinforcement for advanced carbon fiber composites \cite{John,Abd,Abh}.

These outstanding properties have been extensively confirmed by a large bulk of literature aimed at understanding the mechanical behavior of these materials and optimizing their application. 
Wang \emph{et al.} \cite{WaJin13}, for instance, investigated the effect of graphene morphology on the main toughening mechanisms of polymer/graphene nanocomposites. They found that the toughening effect strongly depends on the size of graphene sheets with the best performance being related to the smallest sheet sizes and the main toughening mechanism being micro-crack deflection. 

A comprehensive analysis on the mechanisms of damage of graphene based nanocomposites was carried out by Chandrasekaran \emph{et al.} \cite{ChaSei13, ChaSa14} who showed that micro-crack deflection, crack pinning and separation of graphene layers are the main sources of energy absorption. Further, the thermoset system investigated showed an outstanding enhancement of fracture toughness and electric conductivity.

An improvement of the fracture toughness and thermal conductivity was reported by Lee \emph{et al.} \cite{LeeWan16}, who studied thermoset polymers modified by functionalized graphene, whereas the effect of graphene dispersion on the mechanical properties was investigated by Tang \emph{et al.} \cite{TanWan13}. Several other studies confirm the outstanding performance of graphene nanocomposites (see, among others, \cite{JiKu13,ShoGho14, GalWan14}).

While a large bulk of data on the mechanical properties of graphene nanocomposites are available already, an aspect often overlooked in the literature is the scaling of the fracturing behavior. This is a serious issue since the design of large nanocomposite structures or small-scale graphene-based electronic components requires capturing the scaling of their mechanical properties. This is challenging since, due to the complex mesostructure characterizing graphene nanocomposites (and other quasi-brittle materials such as concrete, ceramics, rocks, sea ice, and many bio-materials, just to mention a few), the size of the non-linear Fracture Process Zone (FPZ) occurring in the presence of a large stress-free crack is usually not negligible \cite{Baz84,Baz90,bazant1996_1,bazant1998_1,salviato2016_1}. The stress field along the FPZ is nonuniform and decreases with crack opening, due to discontinuous cracking, nano-crack deflection by graphene platelets, and frictional pullout of graphene layers \cite{ChaSei13, WaJin13, ChaSa14, ShoGho14}. As a consequence, the fracturing behavior and, most importantly, the energetic size effect associated with the given structural geometry, cannot be described by means of the classical Linear Elastic Fracture Mechanics (LEFM). To capture the effects of a finite, non-negligible FPZ, the introduction of a characteristic (finite) length scale related to the fracture energy and the strength of the material is necessary \cite{Baz84,Baz90,bazant1998_1,bazant1996_1,salviato2016_1}.

This work proposes an investigation on the structural scaling of polymer graphene nanocomposites. Fracture tests on geometrically scaled Single Edge Notch Bending (SENB) specimens with varying contents of graphene were conducted to study the effects of nanomodification on the scaling. It is shown that, while the scaling of the pristine polymer follows Linear Elastic Fracture Mechanics (LEFM), this is not the case for nanocomposites, even for very low graphene contents. Through the analysis of the nominal strength as a function of the specimen size, it is shown that small specimens have a more pronounced ductility with limited scaling and a significant deviation from LEFM whereas larger specimens behave in a more brittle way, with scaling of nominal strength closer to the one predicted by LEFM. This behavior is due to the significant size of the FPZ compared to the specimen size which affects the overall fracturing behavior.

Accounting for the FPZ size is of utmost importance to capture the scaling of the structural behavior and to correctly estimate the fracture energy of the material from fracture tests. It is shown that the use of LEFM to estimate mode I fracture energy leads to non-objective results, the fracture energy depending on the size of the specimen tested. It is also shown that, by introducing a length scale related to the FPZ size by means of an Equivalent Crack approach, a formula for the scaling which depends not only to the material fracture energy but also to the FPZ size can be derived. This formula, known as Ba\v{z}ant's Size Effect Law (SEL) \cite{Baz84,Baz90,bazant1996_1,bazant1998_1}, enables an excellent fitting of the experimental data and the objective estimation of the fracture energy and the FPZ size. A comparison between LEFM and SEL showed that for the small specimen sizes investigated, the difference can be as high as 113\%, this difference decreasing for larger specimens as the effects of the FPZ become less significant in the context of the larger portion of the specimens in the linear elastic regime.
\section{Materials and Methods}
\subsection{Materials and preparation}
\label{materialprep}

The thermoset polymer used for all of the tested specimens was composed by an $\textnormal{EPIKOTE}\textnormal{\texttrademark}$ Resin  $\textnormal{MGS}\textnormal{\texttrademark}$ and an $\textnormal{EPIKOTE}\textnormal{\texttrademark}$ Curing Agent $\textnormal{MGS}\textnormal{\texttrademark}$ RIMH 134-RIMH 137 (Hexion \cite{hexion}) combined in a 100:36 ratio (by weight).  Thanks to the low viscosity of the resin, a solvent was not used even after the addition of graphene.

The nanofiller was A-12 Graphene Nanoplatelet (Graphene Supermarket \cite{graphenesupermarket}) with an average flake thickness of less than 3 nm (between 3-8 graphene monolayers) and lateral dimensions of approximately 2-8 microns.

\begin{figure}[H]
\center
\includegraphics[scale=0.8]{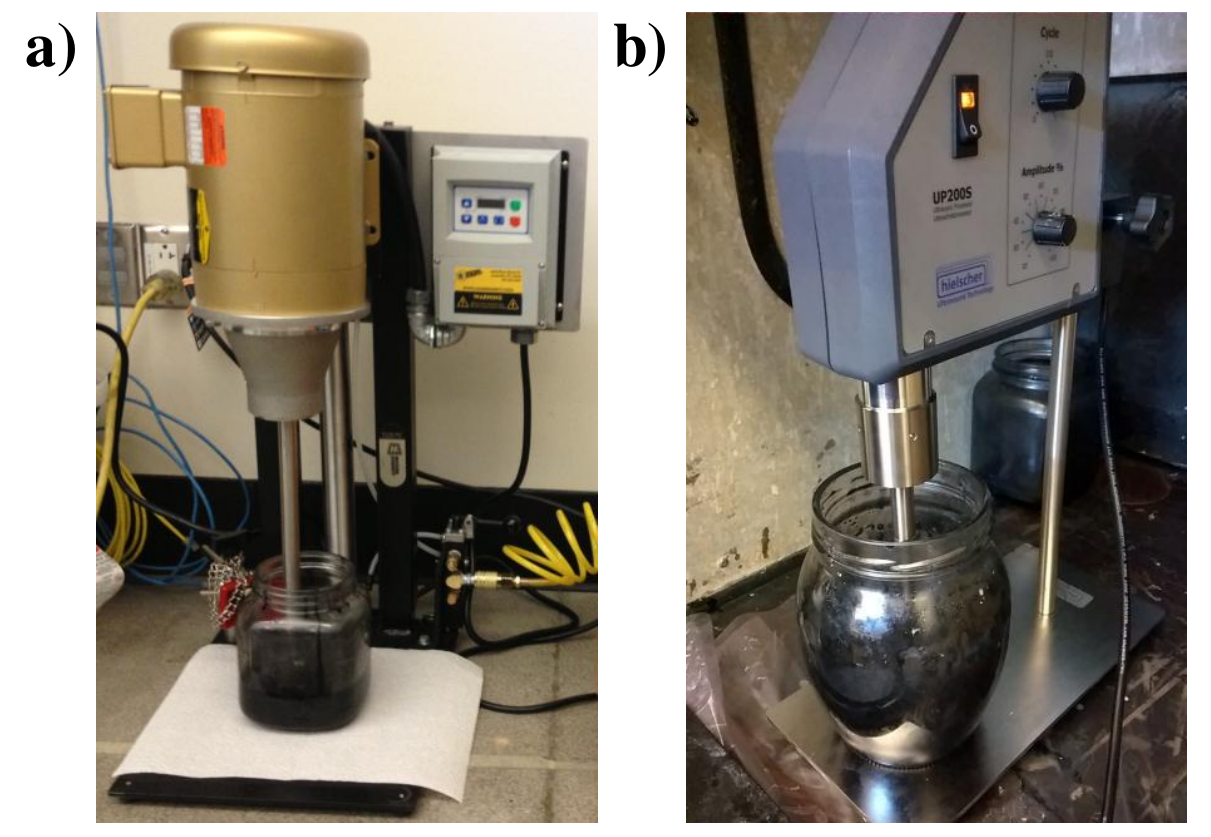}
\caption{Preparation of graphene nanocomposites: a) high shear mixing; b) sonication.}
\label{fig:mixerandsonic}
\end{figure}

The epoxy and hardener were manually mixed for 10 minutes and poured into silicone molds made of RTV silicone from TAP Plastics \cite{tapplastics} to create geometrically scaled specimens with consistent sizes. For the preparation of the epoxy/graphene specimens, the desired amount of epoxy and graphene was mixed for 10 minutes and then, as Figure \ref{fig:mixerandsonic}a shows, high shear mixed at 1500 rpm for 20 minutes by means of an electrically activated high shear mixer with a 48 mm impeller (Mixer Direct \cite{mixer}). Shear mixing was followed by sonication using a Hielscher UP200S sonicator \cite{hielscher} with a 7 mm sonotrode for 20 minutes at 70\% amplitude and a duty cycle of 0.5 (Figure \ref{fig:mixerandsonic}b). This latter step was required to promote a satisfactory platelet exfoliation and it had been adopted successfully by the authors for the dispersion of nanoclays in thermoset polymers \cite{quaresimin2011_1,zappalorto2013_1,zappalorto2013_2}).

In order to remove any air bubbles, the mixture was then degassed for 20 minutes in a vacuum trap using a Robinair 15400 vacuum pump \cite{robinair}.  After degassing and adding the hardener, the mixture was manually mixed for 10 minutes and then poured into the silicone molds. It was allowed to cure at room temperature for approximately 48 hours and then post-cured in an oven for 4 hours at $60$ $^{\circ}$C. 

After curing, the specimens were pre-cracked through a three-stage process.  The first step consisted in creating a notch about one quarter of the specimen width by means of a 0.2 mm wide diamond coated saw. Then, during the second step, the specimens were chilled at about $0$ $^{\circ}$C for approximately 8 hours to facilitate the creation of the crack by tapping.  Tapping was preferred to sawing to create the last portion of the crack in order to provide a very sharp tip and to limit the emergence of plastic residual stresses \cite{xiao,cayard}. Since the epoxy/graphene specimens were completely black opaque, even at the lowest concentration, the identification of the crack tip was very difficult. To overcome this issue, the specimens were painted white so that the contrast in colors provided a better observation of the crack tip location. This was quintessential to guarantee a proper geometrical scaling with all the crack lengths within 0.35-0.55 of the width of the specimens.

\subsection{Specimen preparation}

\subsubsection{Uniaxial testing}
\begin{figure}[H]
\center
\includegraphics[scale=0.5]{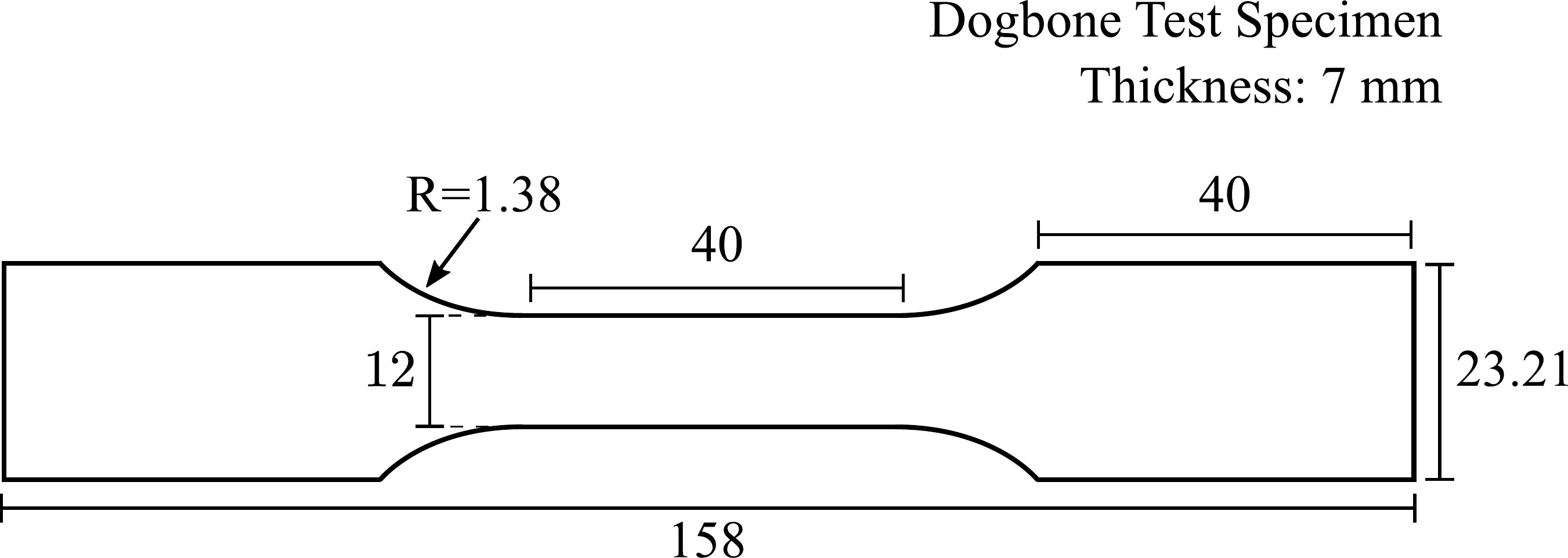}
\caption{Dogbone specimen geometry. Units: mm.}
\label{fig:dogbonedim}
\end{figure}

The dogbone specimens used for the uniaxial tests followed ASTM D638-02a \cite{ASTM_dogbone}.  The specimens, illustrated in Figure \ref{fig:dogbonedim}, were designed to avoid significant stress concentrations which would have caused failure outside the gauge length.  Four material configurations, characterized by different graphene weight contents, were prepared for uniaxial tensile tests, namely: pure epoxy, 0.3 wt$\%$, 0.9 wt$\%$, and 1.6 wt$\%$. The surfaces of dogbone specimens were painted white and then speckled with black paint to allow for Digital Image Correlation (DIC) analysis.  DIC was adopted to provide accurate information on the entire strain field and to analyze possible strain localizations which are known to occur in polymer testing. Thanks to DIC, no mechanical device was attached to the specimens during the experiments. 

\subsubsection{Fracture testing}

The design of the Single Edge Notch Bending (SENB) specimens was based on ASTM D5045-99 \cite{ASTM_SENB}.  Four sets of SENB specimens were prepared for the three-point bending tests: pure epoxy, 0.3 wt$\%$, 0.9 wt$\%$, and 1.6 wt$\%$ graphene nanocomposites.  In order to study the scaling of the fracturing behavior, as illustrated in Figure \ref{fig:specimendimensions}, geometrically scaled specimens of three different sizes were prepared for each material configuration.  The dimensions, scaled as 1:2:4, were 10x36 mm, 20x72 mm, and 40x144 mm, respectively. The various crack lengths of the specimens were approximately in the range 0.35$D$ to 0.55$D$, where $D$ is the width of the specimen.

\subsection{Testing}
The uniaxial tensile tests and three-point bending tests were performed on a closed-loop electro-activated 5585H Instron machine. The speckled dogbone specimens were analyzed by means of a Digital Image Correlation (DIC) system by Correlated Solutions \cite{correlated}, composed of two cameras and a workstation for post-processing, synchronized with the load frame. The load rate for uniaxial tensile tests was 5 mm/min whereas, to avoid viscoelastic effects, the load rate for three-point bending tests was adjusted for the different sizes to achieve roughly the same average strain rate of 0.2 \%/min.  It is worth mentioning here again that the geometrical scaling involved also the length of the initial crack which was always about 0.35-0.55 the width, $D$, of the specimens.  The scaling did not involve the thickness, $t$, which was kept about 12 mm for all the investigated sizes.

\subsection{Microscopic analysis}
Scanning Electron Microscopy (SEM) was performed to investigate the toughening mechanisms of the material system at the nanoscale.  The fracture surfaces were observed by means of a JSM-6010PLUS/LA Electron Microscope \cite{jeol} by applying an acceleration voltage of 1 kV without sputtering.

\section{Experimental Results}
\subsection{Uniaxial tests}

\begin{figure} [H]
\center
\includegraphics[scale=0.5]{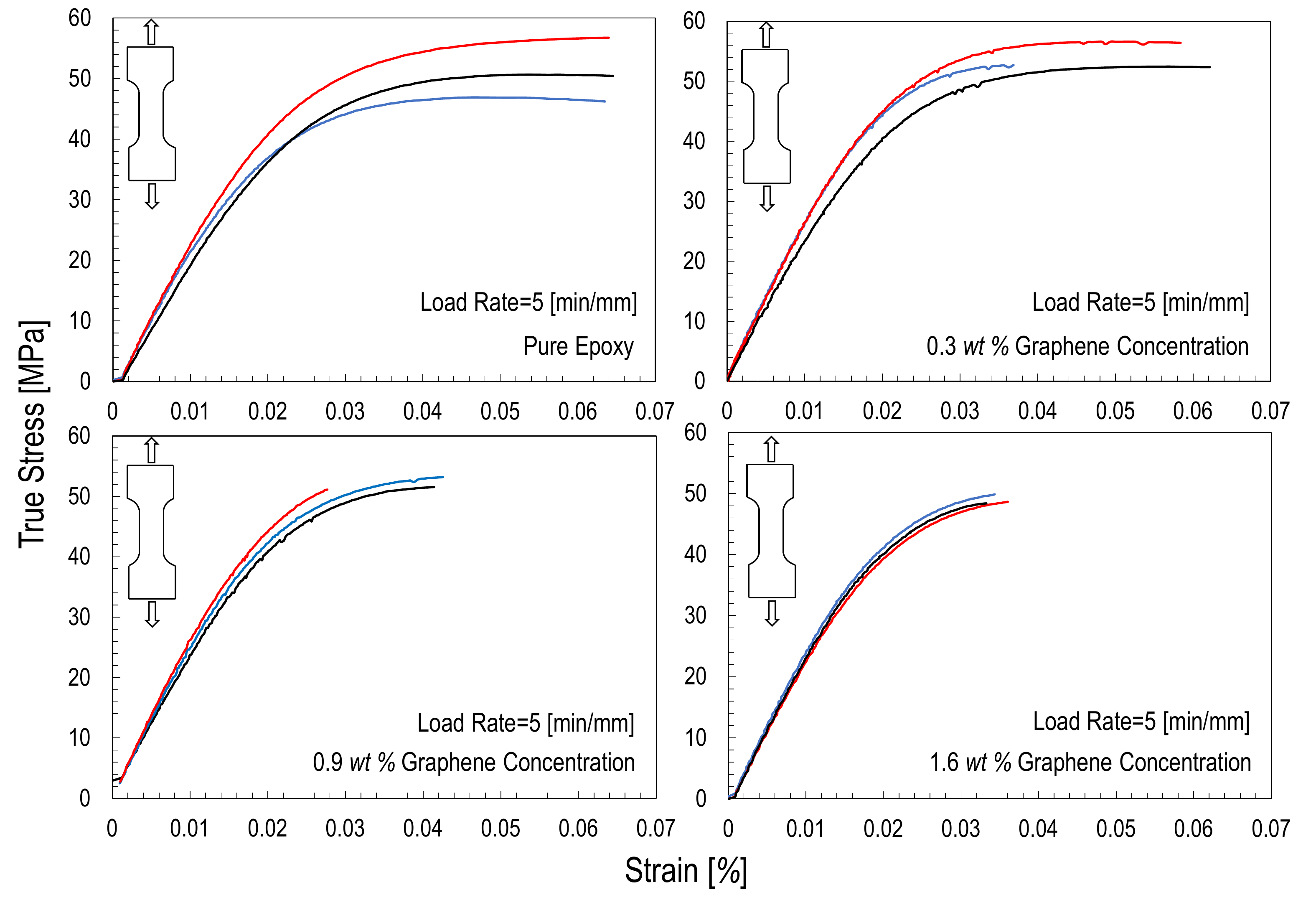}
\caption{True stress vs strain measured from tensile tests on dogbone specimens.}
\label{fig:truestressstrain}
\end{figure}

The true stress and strain curves obtained from uniaxial tensile tests are plotted in Figure \ref{fig:truestressstrain} for dogbone specimens of different graphene concentrations. It is worth mentioning that, thanks to DIC, the stress-strain curves could be characterized in the presence of strain localization in the specimens, enabling the investigation of the behavior of the material at large deformations. This was particularly important since, as can be noted, all the tests were characterized by a significant non-linear behavior which becomes less and less significant with increasing graphene content.  This can be due to a higher presence of voids or defects with higher amounts of graphene which may lead to localizations and failures during the non-linear deformation. 

\begin{figure} [H]
\center
\includegraphics[scale=0.5]{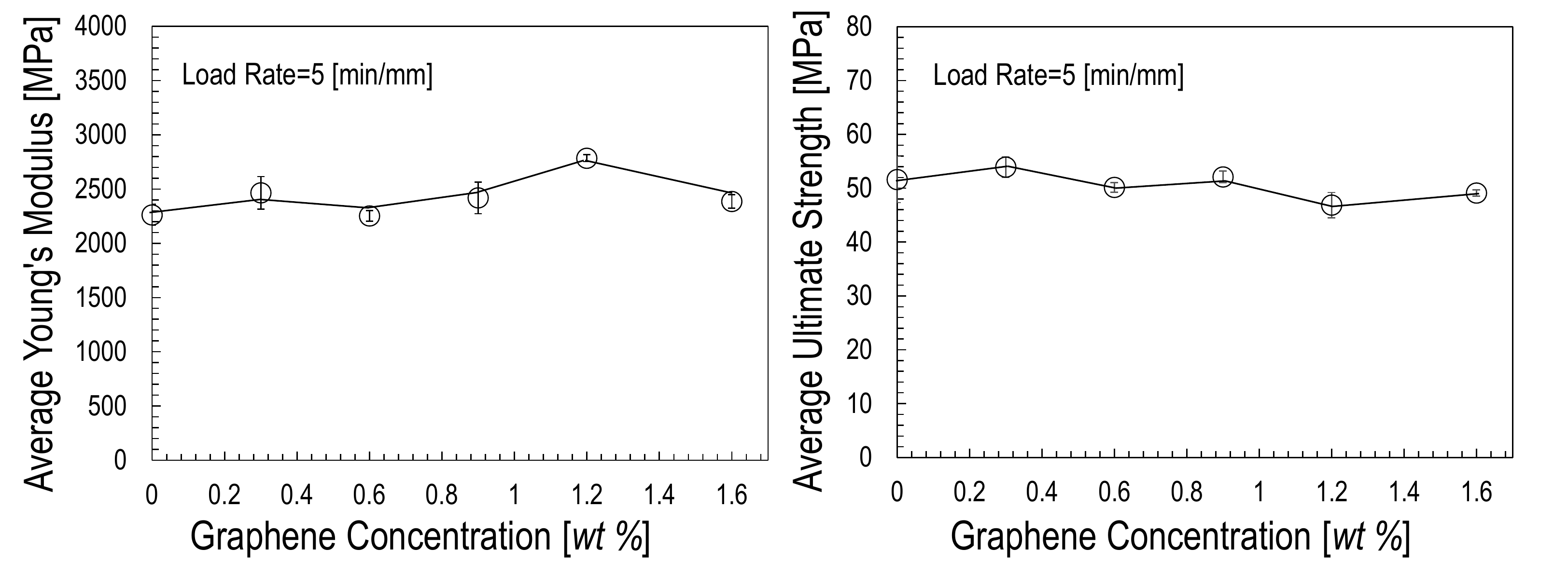}
\caption{Young's modulus and ultimate strength of graphene nanocomposites as a function of graphene content.}
\label{fig:tensileandym}
\end{figure}

The Young's modulus and ultimate tensile strength are plotted in Figure \ref{fig:tensileandym}. As can be noted, the addition of graphene did not significantly affect the elastic behavior and strength of the material investigated. 

\subsection{Fracture tests}

\begin{figure} [H]
\center
\includegraphics[scale=0.5]{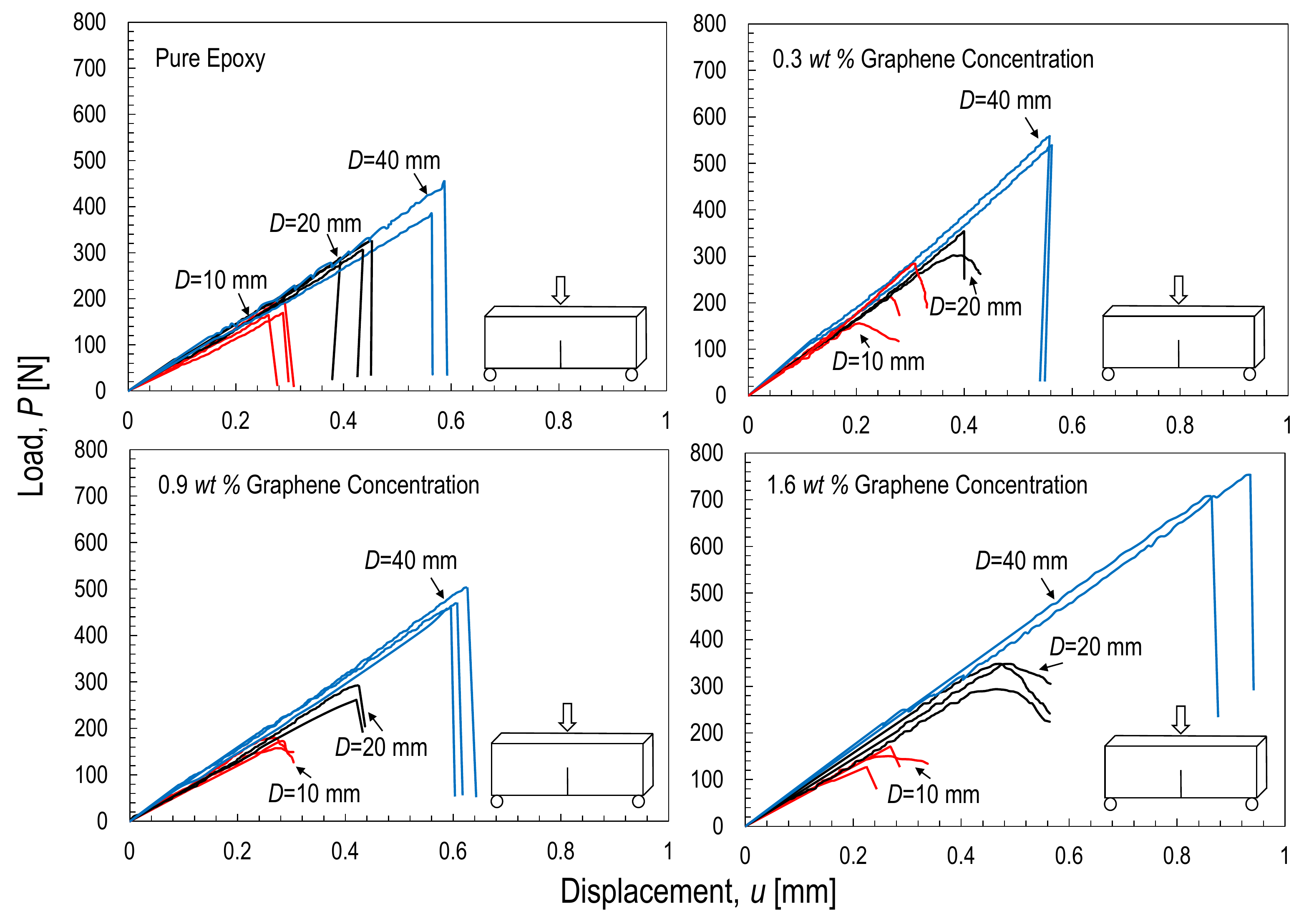}\caption{Load-displacement curves for different graphene concentrations and specimen sizes.}
\label{fig:Loaddispcurve}
\end{figure}

The load-displacement curves of three-point bending tests are plotted in Figure \ref{fig:Loaddispcurve} for different graphene concentrations. It is worth noting that, for the pure epoxy specimens, the mechanical behavior is linear up to the peak load which is followed by unstable crack propagation. This is an indication of pronounced brittle behavior for all the sizes investigated.  With the addition of graphene, differences in the behavior of small specimens with respect to large specimens become visible, this effect being more pronounced for higher graphene contents.  In fact, while large specimens show a very linear response up to failure, a significant non-linear segment before the peak load characterizes the smaller sizes.  This latter aspect indicates hardening inelastic behavior and reduced brittleness (or higher ductility) for the smallest specimen sizes.  After reaching the peak load, the specimens exhibited snap-back instability for all the investigated sizes and graphene concentrations.  As a consequence, the failures were catastrophic (dynamic) and occurred shortly after the peak load. 

The crack length, maximum load, and nominal strength $\sigma_{Nc}=3P_cL/2tD^2$ for geometrically scaled specimens of different Graphene Concentrations (GC) are tabulated in Table \ref{tab:crackloadstress}. In the definition of nominal strength, $P_c$ is the critical load, $t$ is the thickness of the specimens, $L$ is the span between the two supports, and $D$ is the width of the specimens.

\subsection{SEM analysis}
\begin{figure} [H]
\center
\includegraphics[scale=0.5]{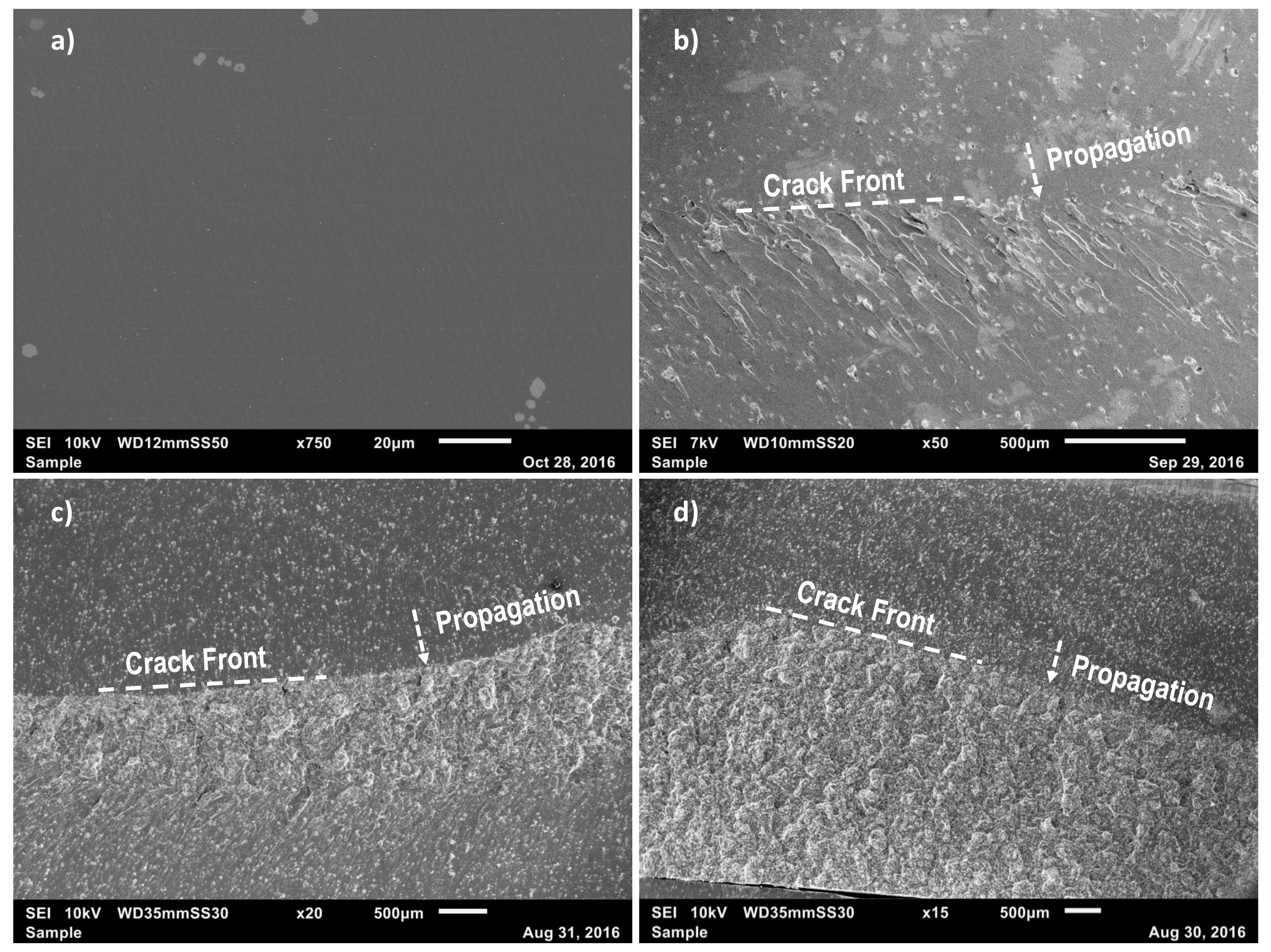}
\caption{Fracture surfaces of SENB specimens for different graphene concentrations: a) Pure epoxy; b) 0.3 wt\%; c) 0.9 wt\%; d) 1.6 wt\%.}
\label{fig:SEMwide}
\end{figure}

In order to investigate the main nanoscale mechanisms of damage that can lead to an increase in mode I fracture energy of graphene nanocomposites, the SENB specimens were cut and the fracture surfaces were gold-coated in order to be used for Scanning Electron Microscopy (SEM) by a JSM-6010PLUS/LA Electron Microscope \cite{jeol}.  The SEM images of some samples are showed in Figure \ref{fig:SEMwide} highlighting the differences between each graphene concentration. As can be noted, the fracture surface of the pure epoxy specimen was very smooth (Figure \ref{fig:SEMwide}a) whereas the surface becomes rougher in texture as the graphene concentration increases (Figure \ref{fig:SEMwide}b-d). 

\begin{figure}[H]
\center
\includegraphics[scale=0.6]{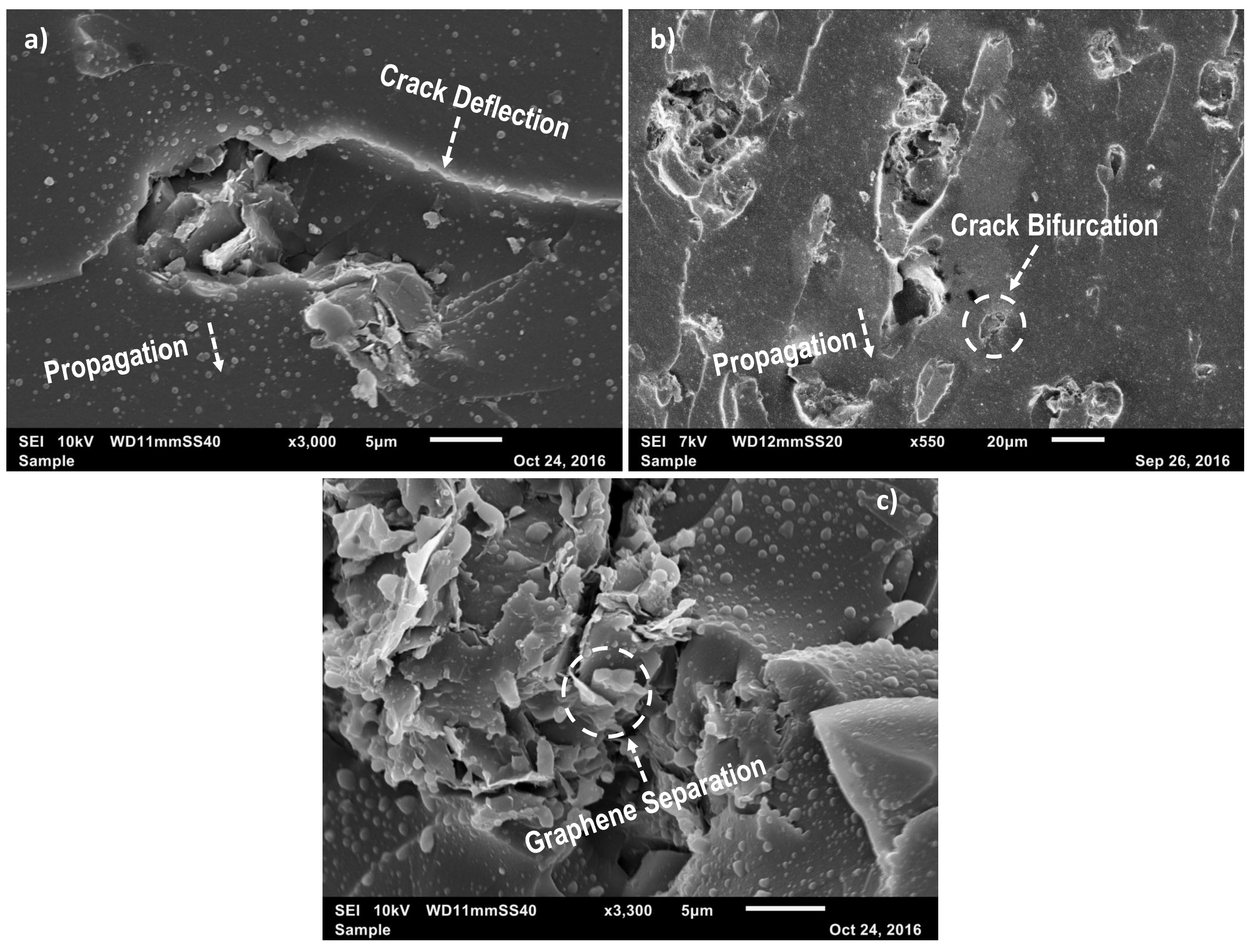}
\caption{Damage mechanisms of graphene nanocomposites: a) crack deflection (1.6 wt\% graphene concentration); b) crack pinning/bifurcation (0.9 wt\% graphene concentration); c) separation between graphene layers (1.6 wt\% graphene concentration).}
\label{fig:SEMdamage}
\end{figure}

Higher magnification images in the propagation region are shown in Figures \ref{fig:SEMdamage}a-c for 0.9 wt$\%$ and 1.6 wt$\%$ graphene concentrations. Based on the pictures, the damage mechanisms are shown to be the following:  a) microcrack deflection; b) microcrack pinning; and c) separation between graphene layers.  As Figure \ref{fig:SEMdamage}a shows, crack deflection occurs when the crack front meets the surface of the graphene sheets and the crack is deflected leading to crack propagation around the graphene sheet into another plane. On the other hand, Figure \ref{fig:SEMdamage}b shows that when the crack front meets the surface of the graphene sheets, it becomes pinned and splits into two cracks. Finally, as shown in Figure \ref{fig:SEMdamage}c, when the crack front meets the edge of the graphene sheets, the crack continues to propagate in between the layers, splitting the agglomerate in two. 

\begin{figure} [H]
\center
\includegraphics[scale=0.5]{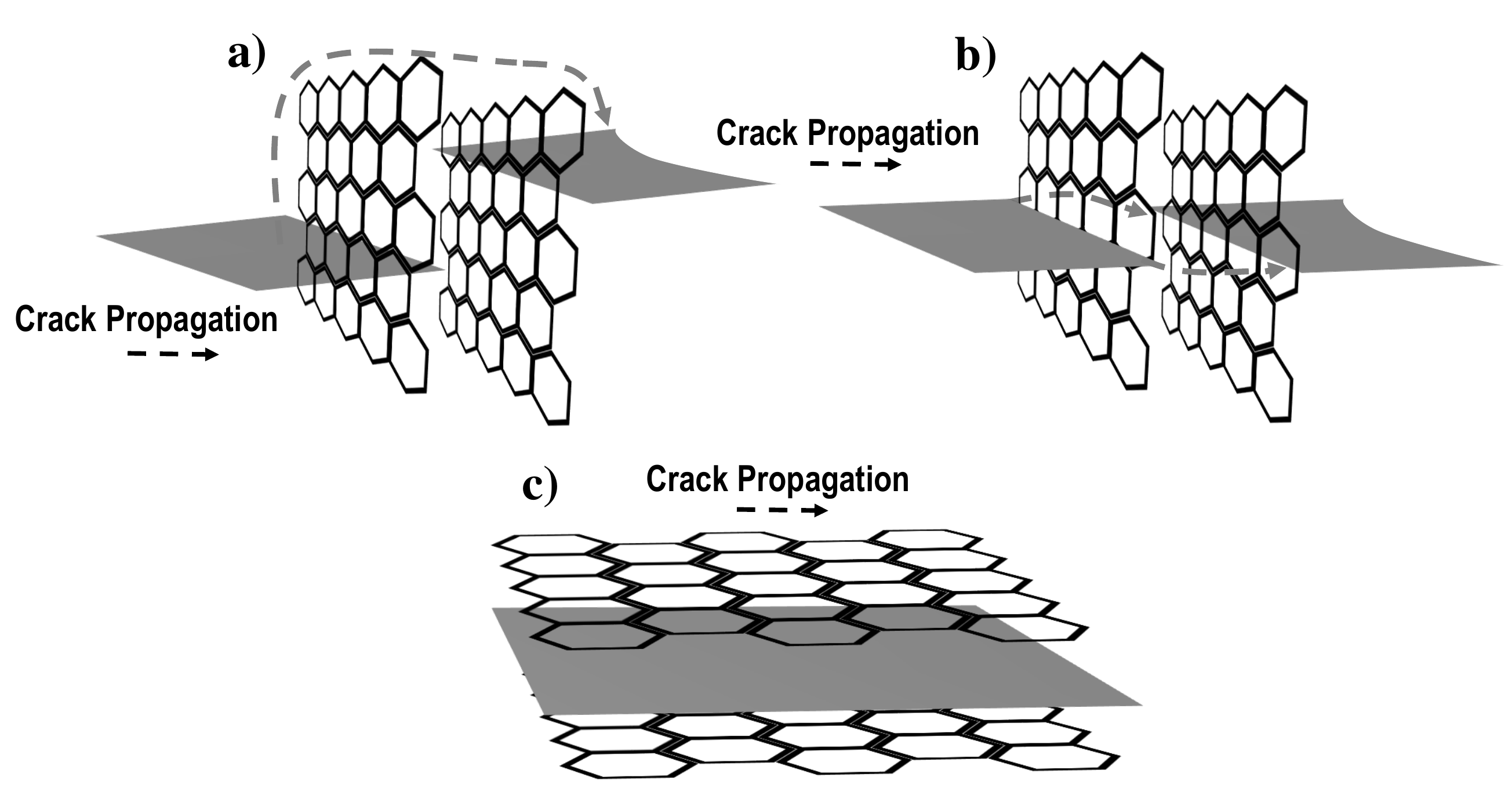}
\caption{Schematic of the main damage mechanisms of graphene nanocomposites reported in this work: a) crack deflection; b) crack bifurcation/pinning; c) separation between graphene layers.}
\label{fig:damageschematic}
\end{figure}

These damage mechanisms, schematically illustrated in Figure \ref{fig:damageschematic}, cause the crack to take a more torturous path thus requiring more energy to be released during the crack propagation. It is worth mentioning that similar damage mechanisms were reported by Chandrasekaran \emph{et al.} \cite{ChaSei13,ChaSa14} for graphene nanocomposites and by Quaresimin \emph{et al.} \cite{quaresimin2011_1} and Zappalorto \emph{et al.} \cite{zappalorto2013_1,zappalorto2013_2}) for nanofillers of similar morphology.

\section{Analysis and Discussion}
\subsection{Analysis of fracture tests by Size Effect Law (SEL)}

The size of the non-linear Fracture Process Zone (FPZ) occurring in the presence of a large stress-free crack is generally not negligible. The stress field along the FPZ is nonuniform and decreases with crack opening, due to discontinuous cracking, micro-crack deflection, micro-crack pinning and graphene layer separations \cite{ChaSei13,ChaSa14}. As a consequence, the fracturing behavior and, most importantly, the energetic size effect associated with the given structural geometry, cannot be described by means of classical Linear Elastic Fracture Mechanics (LEFM). To capture the effects of a finite, non-negligible FPZ, the introduction of a characteristic (finite) length scale related to the fracture energy and the strength of the material is necessary \cite{Baz84,Baz90,bazant1998_1,bazant1996_1,salviato2016_1}. This is done in the following sections.

\subsubsection{Size effect law for graphene nanocomposites}

\begin{figure} [H]
\center
\includegraphics[scale=0.8]{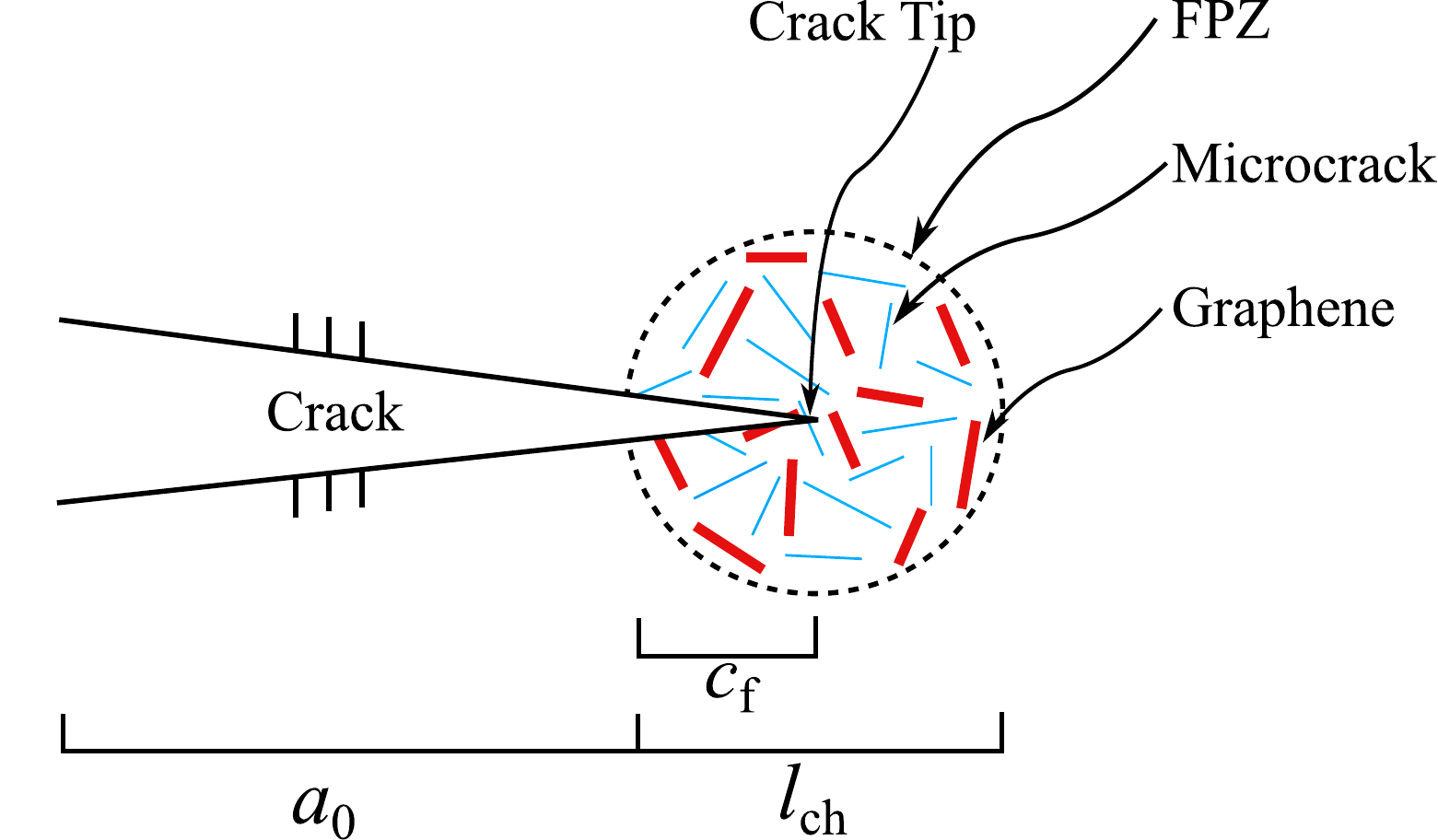}
\caption{Schematic representation of the Fracture Process Zone (FPZ) of graphene nanocomposites and the equivalent crack used for the analysis with $l_{ch}=EG_f/f_t^2=$ Irwin's characteristic length. }
\label{fig:FPZexample}
\end{figure}

The fracture tests can be analyzed leveraging on an equivalent linear elastic fracture mechanics approach to account for the presence of a FPZ of finite size as shown in Figure \ref{fig:FPZexample}. To this end, an effective crack length $a=a_0+c_f$ with $a_0=$ initial crack length and $c_f=$ effective FPZ length is considered. Following LEFM, the energy release rate can be written as follows:
\begin{equation}
G\left(\alpha\right)=\frac{\sigma_N^2D}{E^*}g(\alpha)
\label{eq:Gf}
\end{equation}
where $\alpha=a/D=$ normalized effective crack length, $\sigma_N=3PL/2tD^2=$ nominal stress, $E^*= E$ for plane stress and $E^*= E/\left(1-\nu^2\right)$ for plane strain, and $g\left(\alpha\right)=$ dimensionless energy release rate. The failure condition can now be written as:
\begin{equation}
G\left(\alpha_0+c_f/D\right)=\frac{\sigma_{Nc}^2D}{E^*}g\left(\alpha_0+c_f/D\right)=G_f
\label{failure}
\end{equation}
where $G_f$ is the mode I fracture energy of the material and $c_f$ is the effective FPZ length, both assumed to be material properties. It should be remarked that this equation characterizes the peak load conditions if $g'(\alpha)>0$, i.e. only if the structure has positive geometry \cite{bazant1998_1}.

By approximating $g\left(\alpha\right)$ with its Taylor series expansion at $\alpha_0$ and retaining only up to the linear term of the expansion, one obtains:
\begin{equation}
G_f=\frac{\sigma_{Nc}^2D}{E^*} \left[g(\alpha_0)+\frac{c_f}{D}g'(\alpha_0)\right]
\label{Taylor}
\end{equation}
which can be rearranged as follows \cite{bazant1998_1}:
\begin{equation}
\sigma_{Nc}=\sqrt{\frac{E^*G_f}{Dg(\alpha_0)+c_fg'(\alpha_0)}}
\label{Sel}
\end{equation}
where $g'\left(\alpha_0\right)=\mbox{d}g\left(\alpha_0\right)/\mbox{d}\alpha$.

This equation relates the nominal strength of radially scaled structures to a characteristic size, $D$ and it can be rewritten in the following form:
\begin{equation}
\sigma_{Nc}=\frac{\sigma_{0}}{\sqrt{1+D/D_0}}
\label{eq:sigmaNc2}
\end{equation}
where $\sigma_0=\sqrt{E^*G_f/c_fg'(\alpha_0)}$ and $D_0=c_fg'(\alpha_0)/g(\alpha_0)=$ constant, depending on both FPZ size and specimen geometry. Contrary to classical LEFM, Eq. (\ref{eq:sigmaNc2}) is endowed with a characteristic length scale $D_0$. This is key to describe the transition from ductile to brittle behavior with increasing structure size reported in the fracture tests. 

\subsubsection{Fitting of experimental data by SEL}
The values of $G_f$ and $c_f$ can be determined by regression analysis of the experimental data. Assuming geometrically scaled structures, Ba\u{z}ant \emph{et al}. \cite{bazant1998_1} proposed to define the following:
\begin{equation}
X=D,\quad Y=\sigma_{Nc}^{-2}
\end{equation}
\begin{equation}
\sigma_0=C^{-1/2},\quad D_0=\frac{C}{A}=\frac{1}{A\left(\sigma_0\right)^2}
\end{equation}
thanks to which, Eq. (\ref{eq:sigmaNc2}) can now be expressed in the following linear form:
\begin{equation}
Y=AX+C
\label{eq:slope1}
\end{equation}

\begin{figure}[H]
\includegraphics[scale=0.5]{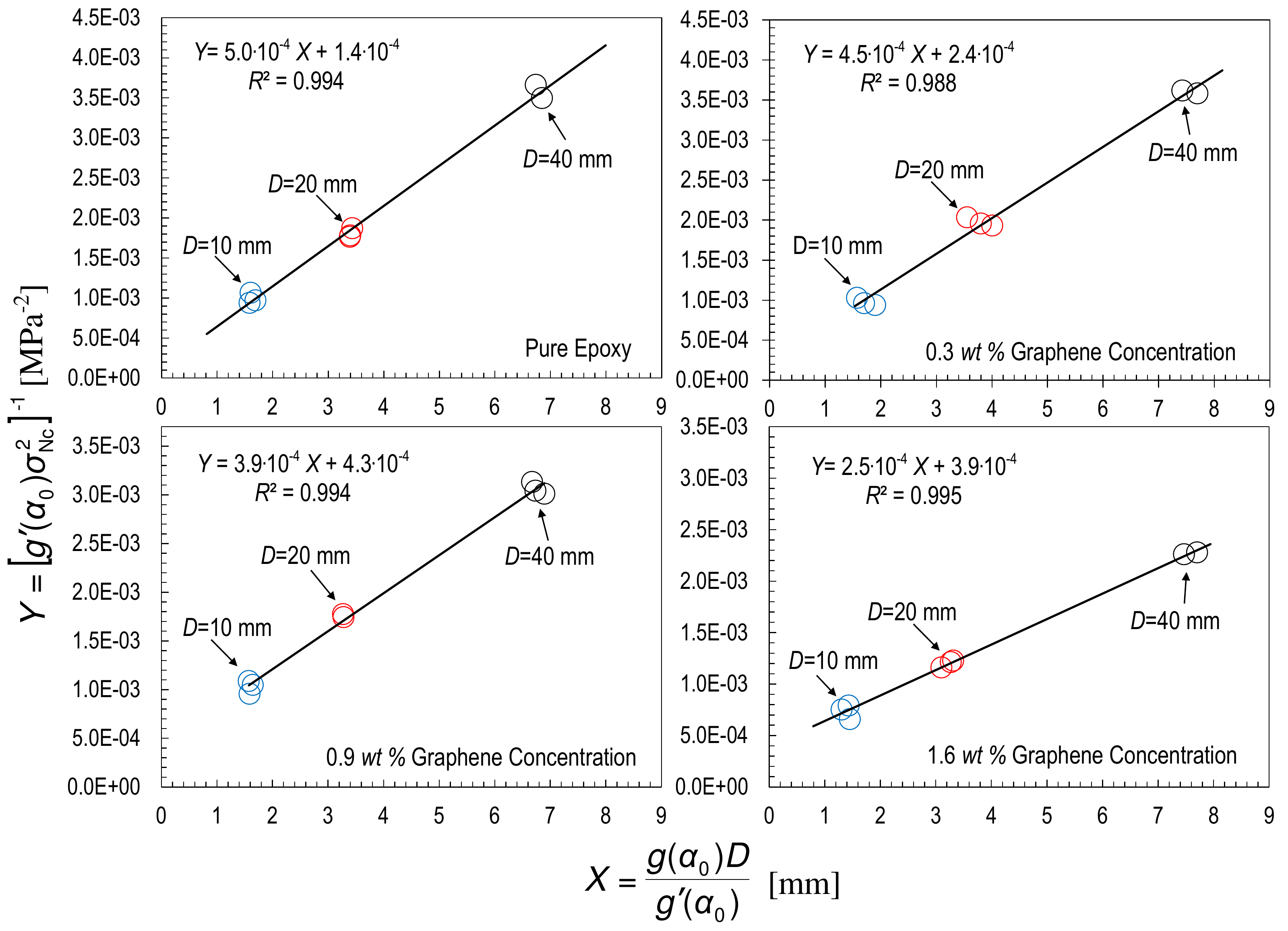}
\caption{Fitting of experimental data through Eq. (\ref{eq:slope2}).}
\label{fig:SELparameters}
\end{figure}

Eq. (\ref{eq:slope1}) can be used to perform a linear regression analysis of the size effect data provided that all the specimens are scaled exactly (i.e. $g(\alpha_0)$ and $g'(\alpha_0)$ take the same values for all the tests). This implies that all the specimens of different sizes have the same normalized crack lengths. However, since the initial cracks for the nanocomposites were created by tapping to avoid residual stresses at the crack tip \cite{xiao,cayard}, a perfect scaling could not be guaranteed. To overcome this issue, in this study, Eq. (\ref{eq:slope1}) was rearranged as follows:
\begin{equation}
\frac{1}{g'(\alpha_0)\sigma_{N_c}^2}=\frac{g(\alpha_0)}{g'(\alpha_0)E^*G_f}D+\frac{c_f}{E^*G_f}
\end{equation}
\begin{equation}
Y=BX+M
\label{eq:slope2}
\end{equation}
where now $X=g(\alpha_0)D/g'(\alpha_0)$, $Y=\left(g'(\alpha_0)\sigma_{Nc}^2\right)^{-1}$, $B=\left(E^*G_f\right)^{-1}$, and $M=c_f/E^*G_f$. Following Eq. (\ref{eq:slope2}), a linear regression analysis was conducted for all the graphene concentrations as represented in Figure \ref{fig:SELparameters} and the parameters $B$ and $M$, reported in Figures \ref{fig:SELparameters}a-d, could be estimated for all the graphene concentrations.

It is interesting to note that the slope $B$ of the regression curve decreases significantly with increasing concentrations of graphene whereas the intercept $M$ shows an opposite trend. As can be noted from the expressions of $B$ and $M$, a milder slope denotes a higher mode I fracture energy thus confirming a toughening effect of the graphene nanoplatelets. Conversely, for increasing fracture energies, a higher value of the intercept indicates an increasing size of the effective Fracture Process Zone (it is worth mentioning here that the case of a material with negligible FPZ, as assumed by LEFM, corresponds to regression lines passing through the origin). This suggests an increasing ductility of the material system with increasing graphene content. 

\subsection{Estimation of fracture properties by SEL}
The parameters of the size effect law, $B$ and $M$, can be directly related to the mode I fracture energy of the material, $G_f$ and the effective FPZ size, $c_f$ as follows:
\begin{equation}
G_f=\frac{1}{E^*B},\quad c_f=\frac{M}{B}
\label{eq:GFCF}
\end{equation}
provided that the functions $g\left(\alpha\right)$ and $g'\left(\alpha\right)=\mbox{d}g\left(\alpha\right)/\mbox{d}\alpha$ and the elastic modulus $E^*$ are known. The mode I fracture energy $G_f$ and the effective FPZ length $c_f$ estimated by means of Eqs. (\ref{eq:GFCF}a,b) are tabulated in Table \ref{tab:Gfandcfvalues} whereas the calculation of $g\left(\alpha\right)$ and $g'\left(\alpha\right)$ is discussed in the next section.

It is worth noting here that, indeed, the addition of graphene led to an enhancement of the mode I fracture energy which increased from $0.880$ N/mm for the pure epoxy case to $1.693$ N/mm for the 1.6 wt$\%$ case, a 92.4\% improvement. On the other hand, the addition of graphene caused a larger FPZ size which changed from $283$ $\mu$m to $1587$ $\mu$m. This latter aspect is of utmost importance: while the inherent assumption of LEFM of a negligible FPZ seems reasonable for the pure epoxy case, this is not true for graphene-modified specimens which show a FPZ about one order of magnitude larger and not negligible compared to e.g. the specimen width. This is in agreement with the results obtained by Salviato \emph{et al.} for other nanocomposite systems by means of multi-scale analytical models \cite{salviato2011_1,salviato2011_2,zappalorto2011_1,zappalorto2011_2,zappalorto2012_1,zappalorto2012_2,salviato2013_1,salviato2013_2,quaresimin2014_1}.

The importance of the finiteness of the FPZ for the estimation of nanocomposite fracture energy and its consequences on the structural scaling will be the subject of section \ref{sec:LEFMSEL}.

\subsection{Calculation of $g(\alpha)$ and $g'(\alpha)$}
\label{sec:gandgprimecalculation}
The function $g(\alpha)$ was obtained through Finite Element Analyses (FEA) in ABAQUS Implicit 6.13 \cite{abaqus}.  8-node biquadratic plain strain quadrilateral elements (CPS8) were adopted while the quarter element technique \cite{barsoum} was used at the crack tip to provide accurate results.  The smallest element size at the tip was about $a_{0}\cdot10^{-5}$ leading to roughly 11,000 elements for the whole model.  A linear elastic isotropic constitutive model was used for the simulations, with $E=3000$ MPa and $\upsilon=0.35$.  The J-integral approach \cite{rice1968_1} was used to estimate the energy release rate in the presence of a concentrated load centered on the top of the SENB specimen.

Once the J-integral was calculated from ABAQUS, the value of $g(\alpha)$ was obtained using the following expression based on LEFM:
\begin{equation}
g(\alpha)=\frac{G(\alpha)E^*}{D\sigma_N^2}
\label{eq:galpha}
\end{equation}
where $\sigma_{N}=3PL/2tD^2$, $P=$ applied load, $L=$ span, $t=$ thickness, $D=$ width, and $\alpha=a/D$ is the normalized effective crack length. 

\begin{figure} [H]
\center
\includegraphics[scale=0.5]{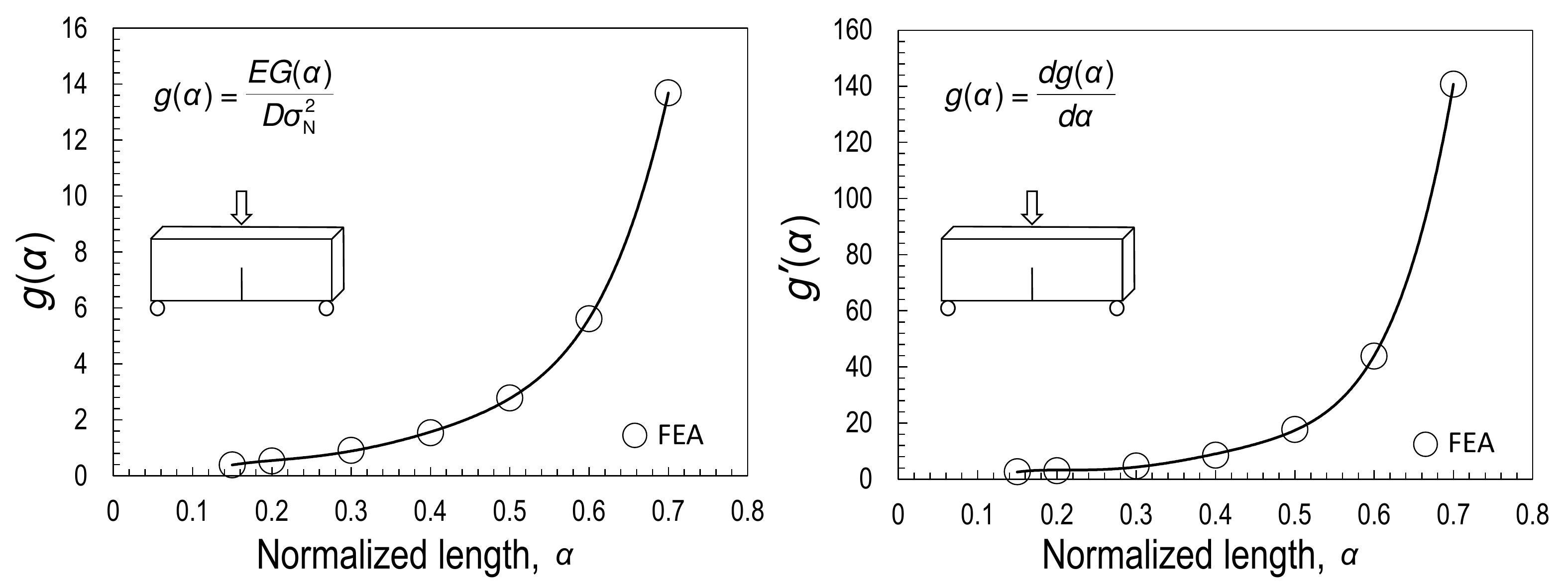}
\caption{Dimensionless energy release rate $g(\alpha)$ and its first derivative $g'(\alpha)$ as a function of normalized crack length $\alpha=a/D$.  The functions are calculated by means of FEA.}
\label{fig:ggprime_interp}
\end{figure}

To determine the function $g'(\alpha)$, various normalized crack lengths close to the selected value of $\alpha$ were considered in order to calculate the tangent slope of $g(\alpha)$ through linear interpolation. Based on the numerical analysis, the following polynomial fitting, plotted in Figure \ref{fig:ggprime_interp}, was obtained:
\begin{equation}
g(\alpha)=1155.4\alpha^5-1896.7\alpha^4+1238.2\alpha^3-383.04\alpha^2+58.55\alpha-3.0796
\end{equation}
\begin{equation}g'(\alpha)=18909\alpha^5-31733\alpha^4+20788\alpha^3-6461.5\alpha^2+955.06\alpha-50.88
\end{equation}

\subsection{Size effect analysis}

\begin{figure} [H]
\center
\includegraphics[scale=0.5]{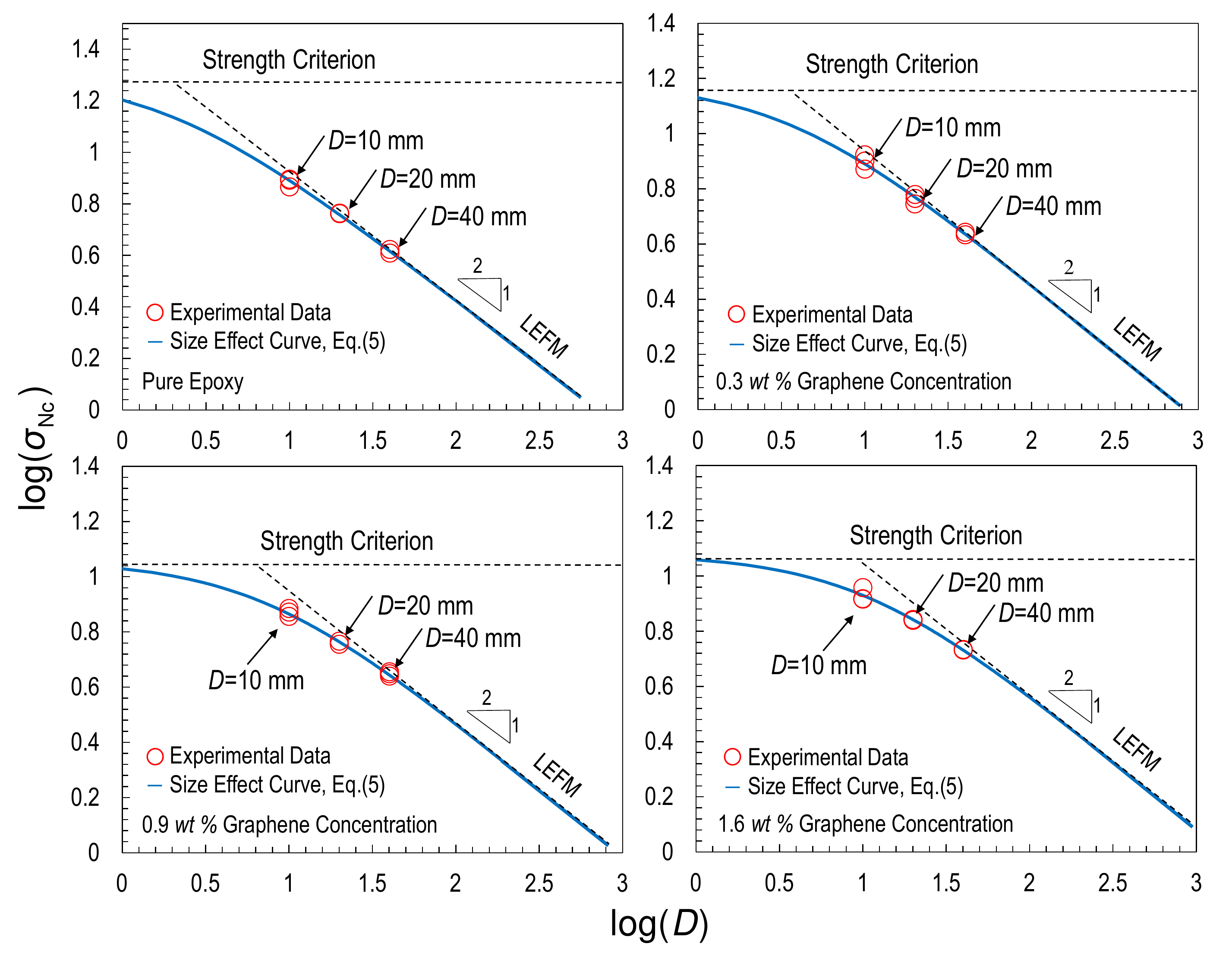}
\caption{Size effect curves for different graphene concentrations.}
\label{fig:sizeeffectcurves}
\end{figure}

To investigate the scaling behavior of structural strength, it is interesting to analyze perfectly radially scaled specimens of different sizes. Now that the fracture properties of the various material systems are known, it is possible to estimate the structural strength for a given normalized crack length starting from the experimental data. In fact, let us consider a crack length of $0.5D$ for all the specimen sizes. One can relate the experimental results, calculated for crack lengths close to $0.5D$ but not exactly $0.5D$, to the desired case by imposing that the energy release rate at failure is $G_f$ in both cases:
\begin{equation}
\displaystyle{\frac{\sigma_{Nc,exp}^{2}D}{E^*}}\left[g(\alpha_{0,exp})+\displaystyle{\frac{c_{f}}{D}}g'(\alpha_{0,exp})\right]=\displaystyle{\frac{\sigma_{Nc,desired}^{2}D}{E^*}}\left[g(0.5)+\displaystyle{\frac{c_{f}}{D}}g'(0.5)\right]
\label{eq:expvsideal}
\end{equation}
This leads to the following expression for the adjusted nominal strength:
\begin{equation}
\sigma_{Nc,desired}=\sigma_{Nc,exp}\sqrt{\frac{Dg(\alpha_{0,exp})+\displaystyle{c_{f}}g'(\alpha_{0,exp})}{Dg(0.5)+\displaystyle{c_{f}}g'(0.5)}}
\label{eq:desiredst}
\end{equation}
The experimental data adjusted according to Eq. (\ref{eq:desiredst}) and the fitting by SEL are shown in Figures \ref{fig:sizeeffectcurves}a-d where the structural strength $\sigma_{Nc}$ is plotted as a function of the structure size $D$ in double logarithmic scale. In such a graph, the structural scaling predicted by LEFM is represented by a line of slope $-1/2$ whereas the case of no scaling, as predicted by stress-based failure criteria, is represented by a horizontal line. The intersection between the LEFM asymptote, typical of brittle behavior, and the plastic asymptote, typical of ductile behavior, corresponds to $D=D_0$, called the \emph{transitional size} \cite{bazant1998_1}.

As can be noted from Figure \ref{fig:sizeeffectcurves}a, the experimental data related to the pure epoxy system all lie very close to the LEFM asymptote showing that, for the range of sizes investigated in this work (or larger sizes), linear elastic fracture mechanics provides a very accurate description of fracture scaling. This confirms that, for the pure epoxy and sufficiently large specimens, the FPZ size has a negligible effect and LEFM can be applied, as suggested by ASTM D5045-99 \cite{ASTM_SENB}. However, this is not the case for graphene nanocomposites which, as Figures \ref{fig:sizeeffectcurves}b-d show, are characterized by a significant deviation from LEFM, the deviation being more pronounced for smaller sizes and higher graphene concentrations. In particular, the figures show a transition of the experimental data from stress-driven failure, characterized by the horizontal asymptote, to energy driven fracture characterized by the $-1/2$ asymptote. This phenomenon can be ascribed to the increased size of the FPZ compared to the structure size which makes the non-linear effects caused by micro-damage in front of the crack tip not negligible. For sufficiently small specimens, the FPZ affects the structural behavior and causes a significant deviation from the scaling predicted by LEFM with a much milder effect of the size on the structural strength. On the other hand, for increasing sizes, the effects of the FPZ become less and less significant thus leading to a stronger size effect closely captured by LEFM. Further, comparing the size effect plots of nanocomposites with different graphene concentrations, it can be noted a gradual shift towards the ductile region thus showing that not only the addition of graphene leads to a higher fracture toughness but also to a gradually more ductile structural behavior for a given size. 

The foregoing conclusions are extremely important for the design of graphene nanocomposite structures or electronic components. As the experimental data show, LEFM does not always provide an accurate method to extrapolate the structural strength of larger structures from lab tests on small-scale specimens, especially if the size of the specimens belonged to the transitional zone. In fact, the use of LEFM in such cases may lead to a significant underestimation of structural strength, thus hindering the full exploitation of graphene nanocomposite fracture properties. This is a severe limitation in several engineering applications such as e.g. aerospace or aeronautics for which structural performance optimization is of utmost importance. On the other hand, LEFM always overestimates significantly the strength when used to predict the structural performance at smaller length-scales. This is a serious issue for the design of e.g. graphene-based MEMS and small electronic components or nanomodified carbon fiber composites in which the inter-fiber distance occupied by the resin is only a few micrometers and it is comparable to the FPZ size. In such cases, SEL or other material models characterized by a characteristic length scale ought to be used. 

\subsection{LEFM vs SEL for the estimation of fracture properties of nanocomposites}
\label{sec:LEFMSEL}
Having discussed the scaling of the fracturing behavior and having shown that, for graphene nanocomposites, the FPZ is not negligible for the range of specimen sizes investigated, it is interesting to check how the mode I fracture energy calculated through SEL compares to the estimation from LEFM for the various sizes and graphene contents.

\begin{figure} [H]
\center
\includegraphics[scale=0.5]{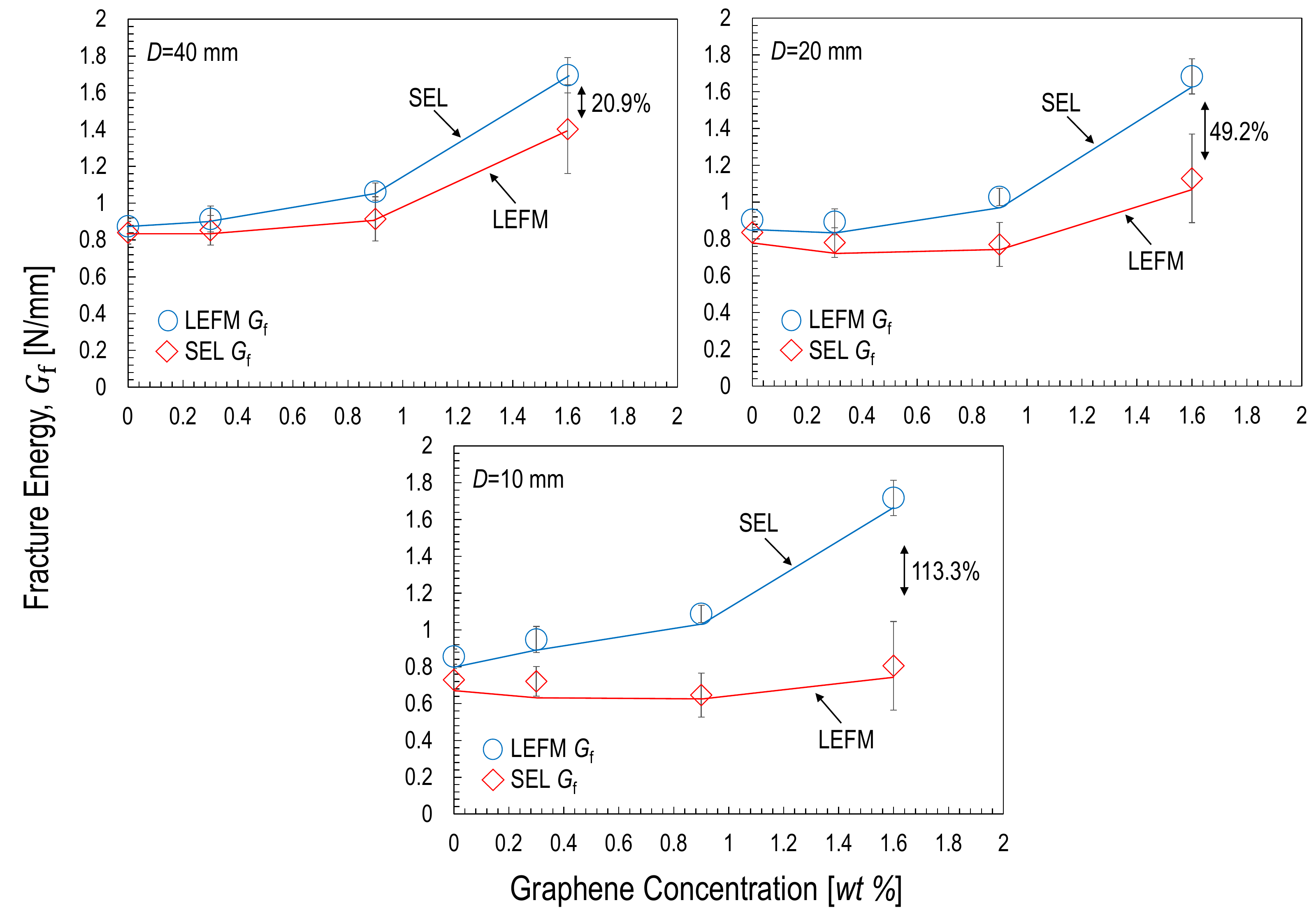}
\caption{Fracture energy estimated from LEFM and SEL for the specimen sizes and graphene contents investigated in this work.}
\label{fig:LEFMvsSEL}
\end{figure}

The fracture energy can be calculated by means of LEFM as follows:
\begin{equation}
G\left(\alpha_0\right)=\frac{\sigma_N^2D}{E^*}g(\alpha_0)
\label{eq:Gf_LEFM}
\end{equation}
where all the quantities and functions have the same meaning discussed in previous sections but, different from SEL, the FPZ length is not accounted for. Figures \ref{fig:LEFMvsSEL}a-c show a comparison between the fracture energy estimated by SEL and by LEFM for different specimen sizes and graphene concentrations. As can be noted, for a given size, the difference between SEL and LEFM increases with the amount of graphene with LEFM underestimating $G_f$. The difference increases with the addition of graphene since, as shown in previous sections, the FPZ size increases monotonically and thus the cardinal assumption of LEFM becomes less and less accurate. The underestimation caused by LEFM can be very significant if one considers that the maximum difference, occurring at $1.6$ wt\% for all sizes, is $20.9\%$, $49.2\%$ and even $113.3\%$ for the large, medium and small sizes respectively. More importantly, a serious issue about using LEFM when the specimen sizes belong to the transitional region is that the estimate is not objective, i.e. it does depend on the size of the specimen tested. This can be noted from Figures \ref{fig:LEFMvsSEL}a-c which show that, for a given graphene content, the fracture energy estimated by LEFM is size dependent, being lower for smaller sizes. It is interesting to note that, for example, the calculations based on LEFM for 1.6 wt$\%$ nanocomposites show basically no increment in the fracture energy for the small size specimen while, for the large size, the increment is about $50\%$. Conversely, thanks to the characteristic length scale associated with the FPZ size, SEL provides the same estimate of the fracture energy for all the sizes. Of course, the size dependence of the fracture energy estimated by LEFM and its difference from SEL depends on the range of sizes investigated: for sufficiently large specimens, both the theories provide the same, size independent, fracture energy. However, as will be shown in a following publication, most of the tests on nanocomposites reported in the literature were performed on small specimens belonging to the transitional region between ductile and brittle behavior. This may explain why the range of fracture energy increments obtained by nanomodification reported in the literature is so large: neglecting the effects of the non-linear FPZ lead to fracture energy estimates which were size dependent and consistently underestimating, the underestimation being more significant for larger nanofiller concentrations and smaller specimen sizes. 

\subsection{Study on the applicability of LEFM to polymer nanocomposites}
The ASTM D5045-99 provides a lower limit for $(D-a)$ to guarantee a plane strain condition.  According to the standard:
\begin{equation}
(D-a)\geq 2.5(K_{IC}/\sigma_y)^2
\label{eq:standard}
\end{equation}
where $D$ is the specimen width, $a$ is the crack length, $K_{IC}$ is the fracture toughness, and $\sigma_y$ is the yielding stress.  Even if this limit is suggested for a different purpose, it is interesting to check if it would be enough to guarantee the use of LEFM to estimate the fracture energy of the material for the graphene nanocomposites investigated in this work.  To this end, Eq. (\ref{eq:standard}) is compared to the width $D_{cr}$ which would be required to guarantee a difference between the nominal strength predicted by LEFM and SEL lower than 10\%.  Accordingly, the lower limit of $D$ can be calculated as follows:
\begin{equation}
\sqrt{\frac{EG_f}{D_{cr}g(\alpha_0)+c_fg'(\alpha_0)}} = 0.9\sqrt{\frac{EG_f}{D_{cr}g(\alpha_0)}}
\label{eq:lefmsel}
\end{equation}
Finally, rearranging Eq. (\ref{eq:lefmsel}), the lower bound $D_{cr}$ can be expressed in the following general form:
\begin{equation}
D_{cr}=4.263\frac{g'(\alpha_0)}{g(\alpha_0)}c_f
\label{eq:Dhalfwidth}
\end{equation}
A comparison between the standard ASTM D5045-99 and SEL for the case in which the crack length is half the width $D$ of the specimen is reported in Table \ref{tab:lefmsel} by using an average Poisson's ratio of $0.4$ for all the graphene concentrations.

According to Table \ref{tab:lefmsel}, the difference between ASTM D5045-99 and size effect curve can be significant depending on the graphene content.  For the highest graphene content and the geometry investigated in this study, $D_{cr}$ (the width providing a difference between LEFM and SEL of $10\%$ only) would be in the order of $40$ mm. This limit is significantly reduced for the pristine epoxy, with LEFM being valid for widths larger than approximately $7$ mm. For both cases, the value calculated from the standard is always lower with $D_{cr}=$ 9.97 mm and 4.45 mm for the 1.6 wt$\%$ and pure epoxy specimens respectively.

\section{Conclusions}
This paper investigated and discussed the effects of nanomodification on the fracturing behavior and scaling of graphene nanocomposites, an aspect of utmost importance for structural design but so far overlooked. The analysis leveraged on a comprehensive set of fracture tests on geometrically scaled Single Edge Notch Bending (SENB) specimens of three different sizes and varying graphene contents. Based on the results obtained in this study, the following conclusions can be elaborated:
\par
\vspace{0.2in}
1. For all the investigated contents, the addition of graphene nanoplatelets to the resin did not provide a significant effect on the elastic properties and strength whereas an outstanding enhancement of mode I fracture energy was reported. A weight fraction of 1.6\% of graphene resulted in an increase of the fracture energy of about 92.4\% compared to the pristine resin, this result making graphene-modified resins ideal candidates for the development of tougher and more ductile composite structures;
\vspace{0.2in}

2. The fracture tests on geometrically scaled SENB specimens confirmed a remarkable size effect. The analysis of the experimental data showed that the fracture scaling of the pure epoxy is captured accurately by Linear Elastic Fracture Mechanics (LEFM). However, this was not the case for graphene nanocomposites which exhibited a more complicated scaling. The double logarithmic plots of the nominal stress as a function of the characteristic size of the specimens showed that the fracturing behavior evolves from ductile to brittle with increasing sizes. For sufficiently large specimens, the data tend to the classical $-1/2$ asymptote predicted by LEFM. However, for smaller sizes, a significant deviation from LEFM was reported with data exhibiting a milder scaling, a behavior associated to a more pronounced ductility. This trend was more and more pronounced for increasing graphene contents;
\vspace{0.2in}

3. The deviation from LEFM reported in the experiments is related to the size of the Fracture Process Zone (FPZ) for increasing contents of graphene. In the pure epoxy the damage/fracture zone close to the crack tip, characterized by significant non-linearity due to subcritical damaging, was generally very small compared to the specimen sizes investigated. This was in agreement with the inherent assumption of LEFM of negligible non-linear effects during the fracturing process. However, the addition of graphene nanoplatelets with the various additional damage mechanisms that come with it (such as e.g. platelet/matrix delamination, nano-crack deflection and plastic yielding), resulted in larger and larger FPZs. For sufficiently small specimens, the size of the highly non-linear FPZ was not negligible compared to the specimen characteristic size thus highly affecting the fracturing behavior, this resulting into a significant deviation from LEFM; 

\vspace{0.2in}

4. Capturing the correct scaling of the fracturing behavior is of utmost importance for structural design. Further, it is quintessential to correctly measure important material properties such as the fracture energy. The analysis of the results reported in this study shows that using LEFM to calculate the mode I fracture energy from the experiments leads to a size dependent $G_f$. Taking the specimens with $1.6\%$ wt of graphene as an example, the fracture energy according to LEFM was 0.8 N/mm, 1.0 N/mm and 1.25 N/mm for the small, medium and large sizes respectively. The reason for this discrepancy is that LEFM lacks intrinsically of a characteristic length and thus cannot capture the effects of the FPZ size;
\vspace{0.2in}

5. Following Ba\v{z}ant \cite{Baz84,Baz90,bazant1998_1}, an Equivalent Fracture Mechanics approach was used to introduce a characteristic length, $c_f$, into the formulation. This length is related to the FPZ size and it is considered a material property as well as $G_f$. The resulting scaling equation, known as Ba\v{z}ant's Size Effect Law (SEL), depends not only on $G_f$ but also on the FPZ size. An excellent agreement with experimental data is shown, with SEL capturing the transition from quasi-ductile to brittle behavior with increasing sizes. The fracture energy for the specimens with $1.6\%$ wt of graphene, finally a material property independent of the specimen size, was $1.69$ N/mm whereas $c_f=$ 1.59 mm;
\vspace{0.2in}

6. The difference between the fracture energy predicted by LEFM and SEL depends on the FPZ size compared to the specimen size, with LEFM underestimating $G_f$ compared to SEL. For the specimens with 1.6$\%$ wt of graphene and the sizes considered in this work, LEFM predictions were about 113.3\%, 49.2\% and 20.9\% lower compared to SEL for the small, medium and large sizes respectively. The difference decreases with increasing specimen sizes and tends to zero for sufficiently large specimens as the FPZ becomes negligible compared to the specimen size.
\vspace{0.2in}

7. The foregoing evidences show that particular care should be devoted to the understanding of the scaling of the fracture behavior of nanocomposites. In particular, the fracture tests carried out to characterize e.g. the fracture energy should guarantee objective results. Size effect testing on geometrically scaled specimens is a simple and effective approach to provide objective data. Alternatively, LEFM could be used provided that the specimen size is large enough. The size limit depends on the size of the FPZ and the geometry of the tested specimen. For the SENB specimens investigated, a difference lower than 10\% between the nominal strength predicted by LEFM and SEL can be guaranteed if $D\ge4.263g'(\alpha_0)c_{f}/g(\alpha_0)$. This limit is generally higher than the one suggested by ASTM D5045-99 to guarantee a plane strain condition.




\section*{Acknowledgments}
Marco Salviato acknowledges the financial support from the Haythornthwaite Foundation through the ASME Haythornthwaite Young Investigator Award. This work was also partially supported by the William E. Boeing Department of Aeronautics and Astronautics as well as the College of Engineering at the University of Washington through Salviato's start up package.

\clearpage
\listoftables
\listoffigures  
\clearpage

\begin{table}[H]
\scriptsize
\centering
\begin{tabular}{ccccc}
\hline
GC (wt\%) &Specimen width (mm) &Crack length (mm) &Max load (N) &Nominal strength (MPa) \\ \hline
0 &D=10 &5.03 &169.33 &7.73\\
0 &D=10 &4.96 &164.62 &7.47\\
0 &D=10 &4.38 &201.01 &9.48\\
0 &D=20 &7.63 &289.27 &7.91\\
0 &D=20 &9.26 &306.58 &6.65\\
0 &D=20 &9.28 &325.62 &6.61\\
0 &D=40 &17.27 &385.44 &4.93\\
0 &D=40 &16.46 &455.31 &5.37\\ \hline
0.3 &D=10 &5.01 &155.00 &7.35\\
0.3 &D=10 &4.04 &212.84 &10.38\\
0.3 &D=10 &3.39 &284.40 &13.45\\
0.3 &D=20 &7.03 &353.95 &9.04\\
0.3 &D=20 &4.88 &538.00 &12.48\\
0.3 &D=20 &7.67 &302.23 &7.78\\
0.3 &D=40 &13.84 &539.04 &6.61\\
0.3 &D=40 &13.13 &558.68 &7.12\\ \hline
0.9 &D=10 &4.62 &179.58 &8.39\\
0.9 &D=10 &5.19 &157.52 &6.87\\
0.9 &D=10 &5.21 &172.87 &7.36\\
0.9 &D=20 &8.46 &292.67 &7.10\\
0.9 &D=20 &9.86 &261.56 &5.94\\
0.9 &D=40 &15.86 &503.51 &6.04\\
0.9 &D=40 &16.93 &469.50 &5.52\\
0.9 &D=40 &17.68 &457.80 &5.15\\ \hline
1.6 &D=10 &5.69 &150.36 &6.51\\
1.6 &D=10 &5.54 &171.04 &7.49\\
1.6 &D=10 &6.02 &126.40 &5.61\\
1.6 &D=20 &9.27 &348.10 &7.76\\
1.6 &D=20 &9.04 &348.27 &7.92\\
1.6 &D=20 &10.35 &294.07 &6.56\\
1.6 &D=40 &13.80 &717.68 &8.41\\
1.6 &D=40 &13.53 &769.07 &8.73\\ \hline
\end{tabular}
\caption{Maximum load and nominal strength reported in SENB tests for different graphene concentrations.}
\label{tab:crackloadstress}
\end{table}

\newpage
\begin{table}[H]
\centering
\scalebox{1}{
\begin{tabular}{ccc}
\hline
Graphene concentration (wt\%) &Fracture energy $G_{f}$ (N/mm)&$c_{f}$ (mm)  \\ \hline
0 &0.880 &0.283\\
0.3 &0.911 &0.546\\
0.9 &1.059 &1.096\\
1.6 &1.693 &1.587\\ \hline
\end{tabular}}
\caption{Experimental values of the mode I fracture energy $G_{f}$ and the effective Fracture Process Zone length $c_{f}$ for different graphene concentrations.}
\label{tab:Gfandcfvalues}
\end{table}

\newpage
\begin{table} [H]
\center
\begin{tabular}{ccc}
\hline
Graphene concentration (wt\%) & ASTM D5045-99 (mm) & Size effect curve (mm)\\ \hline
0 & D $ >4.45$ & D $>7.70$\\
0.3 & D $ >4.60$ & D $>14.80$\\
0.9 & D $ >5.62$ & D $>29.71$\\
1.6 & D $ >9.97$ & D $>43.04$\\ \hline
\end{tabular}
\caption{Lower limit values of the specimen width $D$ according to ASTM D5045-99 and SEL.}
\label{tab:lefmsel}
\end{table}


\newpage
\section*{Figures and Tables}

\newpage

\newpage
\begin{figure} [H]
\center
\includegraphics[scale=0.05]{specimendimensions.pdf}
\caption{Geometry of Single Edge Notch Bending (SENB) Specimens. Units: mm.}
\label{fig:specimendimensions}
\end{figure}


\newpage

\newpage

\newpage

\newpage

\newpage

\newpage

\newpage

\newpage

\newpage

\newpage

\newpage


\end{document}